\documentclass[a4paper,10pt,aps,pre,floatfix,superscriptaddress,nofootinbib]{revtex4}
\usepackage{epsfig,amssymb, color}
\usepackage{bbm, amsmath, graphicx, mathrsfs}
\usepackage{pseudocode}
\usepackage[plain]{algorithm}
\usepackage{hyperref}

\begin{document}

\title{Statistical Mechanics of Budget-Constrained Auctions}
\author{F.~Altarelli}
\affiliation{ISI Foundation, Viale Settimio Severo 65, 10133 Torino, Italy}
\author{A.~Braunstein}
\author{J.~Realpe-Gomez}
\author{R.~Zecchina}
\affiliation{Politecnico di Torino, Corso Duca degli Abbruzzi 24, 10129 Torino, Italy}
\date{April 27, 2009}

\begin{abstract}
Finding the optimal assignment in budget-constrained auctions is a combinatorial optimization problem with many important applications, a notable example being the sale of advertisement space by search engines (in this context the problem is often referred to as the off-line \emph{AdWords} problem). Based on the cavity method of statistical mechanics, we introduce a message passing algorithm that is capable of solving efficiently random instances of the problem extracted from a natural distribution, and we derive from its properties the phase diagram of the problem. As the control parameter (average value of the budgets) is varied, we find two phase transitions delimiting a region in which long-range correlations arise.

\end{abstract}

\maketitle

\section{Introduction}

\subsection{Problem definition}

We consider the following model of budget-constrained auctions, inspired from the sale of advertisement space on internet search engines: a set of \emph{advertisers} $a \in \{1,\cdots,N_\mathrm a\}$ are interested in appearing on the results pages of the searches relative to some \emph{keywords} $k \in \{1,\cdots,N_\mathrm k\}$. Advertiser $a$ offers to pay a \emph{bid} $w_{ka} \in \mathbbm R^+$ in order to appear on the results page each time that keyword $k$ is searched. We assume that all the bids are expressed \emph{before} the auction begins. In this setting, advertiser $a$ doesn't know how many auctions he will end up winning, and therefore how much he will spend. In order to encourage the advertisers to make more bids without risking to spend too much, each advertiser $a$ can also specify a \emph{budget} $B_a \in \mathbbm R^+$ which is the maximum sum that he is willing to pay in a given period of time.

The general problem we want to solve is the following: given the sets of bids $\{w_{ka}\}$ and of budgets $\{B_a\}$, what is the assignment of each keyword $k$ to some advertiser $a$ which maximizes the total revenue for the search engine? We can represent a possible assignment by introducing the binary variable $x_{ka}$ which will take the value 1 if keyword $k$ is assigned to advertiser $a$ and 0 otherwise. The constraint that each keyword be assigned to one and only one advertiser then takes the form
\begin{equation} \label{k_constraint}
  \sum_a x_{ka} = 1 \hspace{0.5cm} (\forall k)
\end{equation}
while the budget constraint takes the form
\begin{equation}
  \sum_k x_{ka} w_{ka} \leq B_a \hspace{0.5cm} (\forall a)
\end{equation}
and the quantity we want to maximize is the revenue of the search engine
\begin{equation}
  \sum_a \sum_k x_{ka} w_{ka}
\end{equation}
where $x \equiv \{x_{ka}\}$. In order to maximize the revenue, we shall make it possible to sell keywords at a discounted price when this allows to saturate the budget of some advertiser which would otherwise remain unsaturated, and define the revenue as
\begin{equation}
  R(x) = \sum_a \min \left[B_a, \; \sum_k x_{ka} w_{ka} \right] \; .
\end{equation}

Clearly, this problem is a variant of the weighted bipartite matching in which the budget constraint has been added. It is also a special case of the general linear Resource Allocation Problem on binary variables, which consists in finding $\{x_i\} \in \{0,1\}^n$ that maximizes $\sum_i a_i x_i$ under the constraints $\sum_j b_{ij} x_j \leq c_i \ (i=1,\cdots,m)$.

\subsection{Summary of known results}

Both the on-line and the off-line versions of the AdWords problem have been the object of considerable attention in recent years. The off-line version is NP-hard even if there are only 2 advertisers \cite{Lehmann01}. An exact algorithm to solve it with time complexity $O(N_\mathrm a \, 4^{N_\mathrm k})$ and an approximate algorithm based on Integer Programming relaxation with an approximation ratio\footnote{The \emph{approximation ratio} is defined as the ratio between the best possible revenue and the revenue achieved by the algorithm in the worst case instance.} $\mathrm e /(\mathrm e -1)$ (on the revenue) are presented in \cite{Andelman04}.
An improvement over this result is provided by \cite{Azar08}, where an algorithm with approximation ratio $3/2$ is introduced.
An approximate algorithm for the on-line version of the problem with competitive ratio\footnote{The \emph{competitive ratio} is defined as the ratio between the best possible revenue and the revenue achieved by the algorithm for the worst case sequence of choices of the on-line instance.} $\mathrm e /(\mathrm e - 1)$ is introduced in \cite{Mehta07}.

These results are concerned with the \emph{worst-case} analysis of the performance (scaling of time and approximation ratio) of the algorithms. In what follows, we shall be interested in a different question: is there an algorithm which has a good performance \emph{on average} for some given ensemble of instances of the problem? The ensemble of instances will be specified by defining the distributions of the bids and of the budgets.

\section{Statistical mechanics analysis of the problem}

\subsection{Factor graph representation and probability marginals} \label{cavity}

We want to consider the problem of budget-constrained auctions as a statistical mechanics system, as done for example in \cite{Galla06}. The configurations $x \equiv \{x_{ka}\}$ will represent the assignments ($x_{ka} = 1$ if keyword $k$ is assigned to advertiser $a$ and 0 otherwise), and the energy function $E(x)$ will represent the portion of the advertisers budgets which remains unspent:
\begin{equation} \label{energy}
  E(x) = \sum_a E_a(x_a) \equiv \sum_a \max \left[0, \, B_a - \sum_{k \in \partial a} w_{ka} x_{ka} \right]
\end{equation}
where $\partial a \equiv \{k | w_{ka} > 0\}$, and $x_a \equiv \{x_{ka} | k \in \partial a\}$. 

We  want to include the constraint (\ref{k_constraint}) that each keyword be assigned to one and only one advertiser as a \emph{hard constraint}. We therefore write the Boltzmann-Gibbs distribution at inverse temperature $\beta$ as:
\begin{equation} \label{boltzmann_gibbs}
  P_\beta(x) = \frac 1 {Z(\beta)} \prod_a \mathrm e^{-\beta E_a(x_a)} \prod_k \mathbbm 1 \left[ \sum_a x_{ka} = 1 \right]
\end{equation}
where the partition function $Z(\beta)$ is a normalization.

This factorization of the probability distribution into local terms involving only some variables and corresponding to the assignment constraints (\ref{k_constraint}) and the energy terms (\ref{energy}) suggests a \emph{factor graph} representation \cite{Kschischang02} in which an instance of the problem is associated to a bipartite graph $\mathcal G = (F, V; E)$ where $F = \{a\} \cup \{k\}$ is the set of \emph{function nodes} corresponding to the constraints, $V = \{x_{ka}\}$ is the set of \emph{variable nodes} and $E$ is the set of edges $(a,x_{ka})$ and $(k,x_{ka})$ such that $w_{ka} > 0$. In the following we shall assume that as $N_\mathrm a, N_\mathrm k \rightarrow \infty$, the average number of bids expressed by each advertiser remains \emph{finite}. This kind of diluted systems can be successfully treated with the \emph{cavity method} of statistical mechanics \cite{Mezard01, Mezard03}, originally applied to optimization in \cite{Mezard02, Braunstein02}.

Given an instance of the problem $\mathcal G = (F, V; E)$, let us consider a \emph{modified} instance $\mathcal G^{(k)}$ in which the function node $k$ is removed, and another instance $\mathcal G^{(a)}$ in which the function node  $a$ is removed (the rest of these graphs being identical to $\mathcal G$). The name \emph{cavity} refers to the missing node in $\mathcal G^{(k)}$ and in $\mathcal G^{(a)}$. We shall now make an assumption which goes under the name of \emph{Replica Symmetric approximation} (RS), and which will be validated \emph{ex-post}: we shall assume that variables $x_{ka}$ and $x_{jb}$ that are \emph{far away} on the graph are uncorrelated, i.e. that their marginal distribution factorizes: $P(x_{ka},x_{jb})=P(x_{ka})P(x_{jb})$. Under this assumption, when a function node is removed from a diluted random factor graph, the variables connected to the removed node become far away from each other and therefore uncorrelated. Let us then consider the marginal distribution $P^{(a)}(x_{ka})$ for variable $x_{ka}$ in the absence of node $a$: the value of $x_{ka}$ is influenced only by the constraint in node $k$, so that
\begin{equation}
  P^{(a)}_{ka}(x_{ka}) = \mathcal N^{(a)}_{ka} \sum_{\{x_{kb} | b \in \partial k \setminus a\}} \mathbbm 1 \left[ x_{ka} + \sum_{b \in \partial k \setminus a} x_{kb} = 1 \right] P \left( \{x_{kb} | b \in \partial k \setminus a\} \right)
\end{equation}
where $\mathcal N^{(a)}_{ka}$ is a normalization factor, $\partial k \equiv \{a | w_{ka} > 0\}$ and the sum is over the $2^{|\partial k| - 1}$ possible configurations of the variables $\{x_{kb} | b \in \partial k \setminus a\}$. In the limit $N,M \rightarrow \infty$, the joint marginal probability $P \left( \{x_{kb} | b \in \partial k \setminus a\} \right)$ appearing on the right hand side can be replaced with the \emph{cavity} joint marginal probability $P^{(k)} \left( \{x_{kb} | b \in \partial k \setminus a\} \right)$ (this corresponds to a self-consistency assumption on the cavity marginal distributions for systems with $(N_\mathrm a, N_\mathrm k)$ function nodes and systems with $(N_\mathrm a-1, N_\mathrm k)$ or $(N_\mathrm a, N_\mathrm k -1)$ function nodes, as explained in detail in \cite{Mezard01, Mezard03}). But in the absence of node $k$, the variables $x_{kb}$ are far away in the graph, and the RS assumption implies that the joint marginal distribution factorises:
\begin{equation} \label{a_marginal}
  P^{(a)}_{ka}(x_{ka}) = \mathcal N^{(a)}_{ka} \sum_{\{x_{kb} | b \in \partial k \setminus a\}} \mathbbm 1 \left[ x_{ka} + \sum_{b \in \partial k \setminus a} x_{kb} = 1 \right] \prod_{b \in \partial k \setminus a} P^{(k)}_{kb}(x_{kb}) \; .
\end{equation}

With the same argument we obtain
\begin{equation} \label{k_marginal}
  P^{(k)}_{ka}(x_{ka}) = \mathcal N^{(k)}_{ka} \sum_{\{x_{ja} | j \in \partial a \setminus k\}} \mathrm e^{-\beta E_a(x_a)} \prod_{j \in \partial a \setminus k} P^{(a)}_{ja}(x_{ja}) \; .
\end{equation}

Since the variables are binary, the cavity marginal distributions have only one parameter and we shall represent them in terms of cavity \emph{fields} $H_{ka}$ and $U_{ka}$:
\begin{equation} \label{fields}
  \frac {P^{(k)}_{ka}(0)} {P^{(k)}_{ka}(1)} = \mathrm e^{-\beta H_{ka}} \; ,  \hspace{1cm}
  \frac {P^{(a)}_{ka}(0)} {P^{(a)}_{ka}(1)} = \mathrm e^{-\beta U_{ka}} \; .
\end{equation}
In terms of the cavity fields, the self-consistent equations for the cavity marginals (\ref{a_marginal}, \ref{k_marginal}) become:
\begin{eqnarray}
  \mathrm e^{-\beta U_{ka}} &=& \frac {\sum_{b \in \partial k \setminus a} P^{(k)}_{kb}(1) \prod_{c \in \partial k \setminus \{a,b\}} P^{(k)}_{kc}(0)} {\prod_{b \in \partial k \setminus a} P^{(k)}_{kb}(0)} \nonumber \\
  &=& \sum_{b \in \partial k \setminus a} \mathrm e^{\beta H_{kb}} \; , \label{U_update} \\
  \mathrm e^{-\beta H_{ka}} &=& \frac { \sum_{x_{a \setminus k}} \mathrm e^{-\beta E_a(0,x_{a \setminus k})} \prod_{j \in \partial a \setminus k} P^{(a)}_{ja}(x_{ja}) } { \sum_{ x_{a \setminus k} } \mathrm e^{-\beta E_a(1,x_{a \setminus k})} \prod_{j \in \partial a \setminus k} P^{(a)}_{ja}(x_{ja}) } \nonumber \\
  &=& \frac { \sum_{ x_{a \setminus k} } \mathrm e^{-\beta \left[ E_a(0,x_{a \setminus k}) - x_{a \setminus k} \cdot U_{a \setminus k} \right] } } { \sum_{ x_{a \setminus k} } \mathrm e^{-\beta \left[ E_a(1,x_{a \setminus k}) - x_{a \setminus k} \cdot U_{a \setminus k} \right] } } \label{H_update}
\end{eqnarray}
where $x_{a \setminus k} \equiv \{x_{ja} | j \in \partial a \setminus k\}$, $U_{a \setminus k} \equiv \{U_{ja} | j \in \partial a \setminus k\}$ and $x_{a \setminus k} \cdot U_{a \setminus k} \equiv \sum_{j \in \partial a \setminus k} x_{ja} U_{ja}$, and where
\begin{equation}
  E_a(x_{ka},x_{a \setminus k }) = \max \left[ 0, \; B_a - x_{ka} w_{ka} - \sum_{j \in \partial a \setminus k} x_{ja} w_{ja} \right] \; .
\end{equation}

We can solve the system of coupled equations (\ref{U_update}) and (\ref{H_update}) by iteration: we start with some random values for the cavity fields $\{H^0_{ka}\}$ and compute $\{U^1_{ka}\}$ using (\ref{U_update}) with the values of $\{H^0_{ka}\}$; then we compute $\{H^1_{ka}\}$ using (\ref{H_update}) and the values of $\{U^1_{ka}\}$; we iterate this procedure until the values $\{U^t_{ka}\}$ and $\{H^t_{ka}\}$ are a fixed point of the system (\ref{U_update}) and (\ref{H_update}) within some prescribed numerical accuracy.

\subsection{Zero temperature limit of the probability marginals}

As we mentioned before, the optimal assignments of a given instance of the problem will correspond to the configurations with lowest energy. In order to select them, we shall take the zero temperature limit $\beta \rightarrow \infty$. When doing this, we must keep in mind that in case there are multiple optimal assignments, some of the variables will take a unique value in all of them, and in physical terms they will be completely ``polarized'', while other variables will be able to take different values in different optimal assignments. In order to take this into account, we are going to assume that as $\beta \rightarrow \infty$ the cavity fields scale as
\begin{equation} \label{scaling}
  H_{ka} = h_{ka} + \beta^{-1} \xi_{ka} + \mathrm o(\beta^{-1}), \hspace{1cm}
  U_{ka} = u_{ka} + \beta^{-1} \eta_{ka} + \mathrm o(\beta^{-1})
\end{equation}
where $h_{ka}$ and $u_{ka}$ are known as \emph{hard} fields and $\xi_{ka}$ and $\eta_{ka}$ are known as \emph{soft} fields. The soft fields become crucial for those variables for which $h_{ka} + u_{ka} = 0$ (these variables are precisely those that are not fully polarized), and they are irrelevant for the other variables.

Assuming the scaling (\ref{scaling}) and substituting it into the equations for the cavity fields (\ref{U_update}) and (\ref{H_update}) we obtain, by comparing the terms of $\mathrm O(1)$ and those of $\mathrm O(\beta^{-1})$:
\begin{eqnarray}
      u_{ka} &=& - \max_{b \in \partial k \setminus a} h_{kb} \; , \label{u_update} \\
      \eta_{ka} &=& - \log \left[ \sum_{\{b \in \partial k \setminus a | \; h_{kb} \, = \, -u_{ka}\}} \mathrm e^{\xi_{kb}} \right] \label{eta_update}
\end{eqnarray}
and
\begin{eqnarray}
  h_{ka} &=& g^*(B_a) - g^*(B_a - w_{ka}) \; , \label{h_update} \\
  \xi_{ka} &=& \log \left[ {\sum_{x_{a \setminus k} \in \{ x_{a \setminus k}^*(B_a - w_{ka}) \}}} \; \mathrm e^{x_{a \setminus k} \cdot \eta_{a \setminus k}} \right] - \log \left[ {\sum_{x_{a \setminus k} \in \{ x_{a \setminus k}^*(B_a) \} }} \; \mathrm e^{x_{a \setminus k} \cdot \eta_{a \setminus k}} \right] \label{xi_update}
\end{eqnarray}
where
\begin{eqnarray}
  g^*(z) &\equiv& \min_{x_{a \setminus k}} g(z, x_{a \setminus k}) \; , \label{g_star}\\
  g(z,x_{a \setminus k}) &\equiv& \max \left[ 0, \; z - x_{a \setminus k} \cdot w_{a \setminus k} \right] - x_{a \setminus k} \cdot u_{a \setminus k} \label{g}
\end{eqnarray}
are functions of the hard cavity fields $\{u_{ja}\}$ even though we don't write explicitly this dependency, and where the two sums in (\ref{xi_update}) are restricted to local configurations $x_{a \setminus k}^*(B_a - w_{ka})$ such that $g(B_a - w_{ka},x_{a \setminus k}^*(B_a-w_{ka})) = g^*(B_a - w_{ka})$ in the first sum and $x_{a \setminus k}^*(B_a)$ such that $g(B_a, x_{a \setminus k}^*(B_a)) \, = \, g^*(B_a)$ in the second sum.

It is possible to give a simple physical interpretation of the hard and soft fields by considering the following quantities:
\begin{eqnarray}
  \varepsilon^{(a)}_{ka}(x_{ka}) &=& \frac {\sum_{\{x_{jb} | (jb) \neq (ka)\}} E^{(a)}(x) \mathrm e^{-\beta E^{(a)}(x)}} {\sum_{\{x_{jb} | (jb) \neq (ka)\}}\mathrm e^{-\beta E^{(a)}(x)}} \; , \\
  n^{(a)}_{ka}(x_{ka}) &=& \mathrm e^{\beta \varepsilon^{(a)}_{ka}(x_{ka})} \sum_{\{x_{jb} | (jb) \neq (ka)\}}\mathrm e^{-\beta E^{(a)}(x)}
\end{eqnarray}
where $E^{(a)}(x)$ is the energy of configuration $x$ on the cavity graph $\mathcal G^{(a)}$.
Clearly, $\varepsilon^{(a)}_{ka}(x_{ka})$ is the average value of the energy (at temperature $\beta$) on the cavity graph $\mathcal G^{(a)}$ when the variable $x_{ka}$ is constrained to take a given value, while $\log(n^{(a)}_{ka}(x_{ka}))$ is the value of the entropy under the same conditions. When $\beta \rightarrow \infty$, $\varepsilon^{(a)}_{ka}(x_{ka})$ tends to the minimum value of the cavity energy over the configurations with given $x_{ka}$, while $n^{(a)}_{ka}(x_{ka})$ tends to the number of configurations that realize this minimum.

The cavity bias for variable $x_{ka}$ can be written as
\begin{eqnarray}
  \mathrm e ^ {-\beta H_{ka}} &=& \frac {P^{(a)}_{ka}(0)} {P^{(a)}_{ka}(1)} = \frac {\sum_{\{x_{jb} | (jb) \neq (ka)\}}\mathrm e^{-\beta E^{(a)}(x_{ka}=0, \{x_{jb}\})}} {\sum_{\{x_{jb} | (jb) \neq (ka)\}}\mathrm e^{-\beta E^{(a)}(x_{ka}=1, \{x_{jb}\})}} = \frac {\mathrm e^{-\beta \varepsilon^{(a)}_{ka}(0)} n^{(a)}_{ka}(0)} {\mathrm e^{-\beta \varepsilon^{(a)}_{ka}(1)} n^{(a)}_{ka}(1)}
\end{eqnarray}
from which we identify
\begin{eqnarray}
  h_{ka} &=& \varepsilon_{ka}^{(a)}(0) - \varepsilon_{ka}^{(a)}(1) \; , \\
  \xi_{ka} &=& \log n_{ka}^{(a)}(1) - \log n_{ka}^{(a)}(0) \; .
\end{eqnarray}
The hard field $h_{ka}$ is then equal to the difference in the average cavity energies of configurations with $x_{ka} = 0$ relative to configurations with $x_{ka} = 1$, while the soft field $\xi_{ka}$ is equal to (minus) the difference in the cavity entropies between the same configurations. The fields $u_{ka}$ and $\eta_{ka}$ associated to a keyword cavity $(k)$ can be interpreted in the same manner.

A few remarks are in order. First, notice that the equations (\ref{u_update},\ref{h_update}) for the hard fields $\{h_{ka}\}$ and $\{u_{ka}\}$ are \emph{independent} of the soft fields $\{\xi_{ka}\}$ and $\{\eta_{ka}\}$.
One can then split the computation into two parts: first, the fixed point of the recursion (\ref{u_update}, \ref{h_update}) is found for the hard fields, giving $\{h^*_{ka}\}$ and $\{u^*_{ka}\}$; then, these values are substituted (as constants) into the recursion (\ref{eta_update}, \ref{xi_update}), which is solved. This is usually faster and simpler than doing a parallel update.

The second remark concerns the absolute values and the signs of the cavity fields: it is easily seen that $g^*(z)$ is a continuous non-decreasing function of $z$, so that the fields $h_{ka}$ are positive reals\footnote{Notice that this contrasts many other combinatorial optimization problems in which the energy, and therefore the hard fields, are discrete quantities.} or zero; this in turn implies that the fields $u_{ka}$ are negative reals or zero. In a way, this could be expected from the message-passing interpretation of the fields: the messages $u_{ka}$ ``express'' the constraint that keyword $k$ be assigned to only one advertiser, and therefore tend to bias the variables towards 0 (corresponding to a negative field), while the messages $h_{ka}$ ``express'' the requirement that the energy be as low as possible compatibly with the budget constraint, and therefore tend to bias the variables towards 1 (corresponding to a positive field). Similarly, it can be seen that the messages $\xi_{ka}$ must be positive or null, so that the messages $\eta_{ka}$ must be negative or null (the argument however is less trivial and we shall omit it).

As a final remark, let us show that all these update rules can be computed efficiently. This is obvious for the equations (\ref{u_update}) and (\ref{eta_update}), which involve finding the minimum among $c-1$ numbers (where $c$ is the connectivity of the keyword considered). By looking for the two largest values of $h_{ka}$ for a given $k$, one can update all the $u_{ka}$ and $\eta_{ka}$ messages on the edges incident on $k$ in linear time in $c$. For the equations (\ref{h_update}) and (\ref{xi_update}) it is less obvious: these equations require finding the minimum over the local configurations $x_{a \setminus k}$ of a function $g(z,x_{a \setminus k})$, i.e. to compute it in $2^{c-1}$ points, where $c$ can be very large as $N_\mathrm a, N_\mathrm k \to \infty$. However, this can also be computed efficiently. Let us consider the functions $g^*_n(z)$ obtained by minimizing $g(z,\{x_1,\cdots,x_n\})$ over the first 
$n$ variables, which we denote by $x_i$ for simplicity. The functions $g^*_n(z)$ obey the recursion:
\begin{eqnarray}
  g^*_{n+1}(z) &\equiv& \min_{\{x_1,\cdots,x_{n+1}\}} \left\{ \max \left[0, \; z - \sum_{i=1}^{n+1} x_i w_i \right] - \sum_{i=1}^{n+1} x_i u_i \right\} \\
  &=& \min \left[ g^*_n(z), \; g^*_n(z - w_{n+1}) - u_{n+1} \right] \label{recursion_g_star}
\end{eqnarray}
with the initial condition
\begin{equation}
  g^*_0(z) = \max[0, \; z] \; .
\end{equation}
It is then possible to compute the function $g^*(z)$ by iterating the recursion (\ref{recursion_g_star}). This can be done either approximately, discretizing the values of $z$, or exactly, by exploiting the fact that $g^*(z)$ is a function which has slope equal to either one or zero in any point, and that the number of points where it changes slope remains finite as the connectivity increases (this actually depends on the distribution of the values of the bids $\{w_{ka}\}$ and of the fields $\{u_{ka}\}$, but it is true for the distributions we shall consider). The sums appearing in the logarithms of (\ref{xi_update}) can be computed together with the function $g^*(z)$ in a similar way.

\subsection{Average energy and entropy in terms of the fields}

For a given instance, the average energy and the entropy at zero temperature can be computed as functions of the fields $\{h_{ka}\},\{\xi_{ka}\}$ (or equivalently $\{u_{ka}\},\{\eta_{ka}\}$) by writing in their definitions,
\begin{eqnarray}
  E &=& \sum_{x} P(x) \, E(x) \; , \\
  S &=& - \sum_{x} P(x) \, \log P(x)
\end{eqnarray}
the distribution $P(x)$ in terms of the values of the fields at the fixed point:
\begin{eqnarray}
  P(x) = \prod_a P_a(x_a) \, \prod_k P_k(x_k) \, \prod_{(ka)} P_{ka}(x_{ka})^{-1} \; ,
\end{eqnarray}
where as before $x_a \equiv \{ x_{ka} | k \in \partial a \}$, $x_k \equiv \{ x_{ka} | a \in \partial k \}$, and
\begin{eqnarray}
  P_a(x_a) &=& \mathcal N_a^{-1} \, \exp \left\{-\beta \left[E_a(x_a) - x_a \cdot U_a \right] \right\} \; , \\
  P_k(x_k) &=& \mathcal N_k^{-1} \, \exp \left\{\beta \; x_k \cdot H_k \right\} \, \mathbbm 1 \left[ \sum_{b \in \partial k} x_{kb} = 1 \right] \; , \\
  P_{ka}(x_{ka}) &=& \mathcal N_{ka}^{-1} \, \exp \left\{ \beta \, x_{ka} (H_{ka} + U_{ka}) \right\}
\end{eqnarray}
and where $\mathcal N_a$, $\mathcal N_k$ and $\mathcal N_{ka}$ are normalizations.

From these equations one obtains the average energy in the limit $\beta \to \infty$:
\begin{equation}
  E = \sum_a \frac {\sum_{x_a \in \{x^*_a\}} \max \left[ 0, \ B_a - x_a \cdot w_a \right] \mathrm e^{x_a \cdot \eta_a} } { \sum_{x_a \in \{x^*_a\}} \mathrm e^{x_a \cdot \eta_a} } \label{av_energy}
\end{equation}
with the same notations as in the previous paragraph: $u_a \equiv \{u_{ka} | k \in \partial a\}$ and similarly for $\eta_a$ and $w_a$;  $x_a \cdot \eta_a \equiv \sum_{k \in \partial a} x_{ka} \eta_{ka}$ and similarly for $x_a \cdot u_a$ and $x_a \cdot w_a$; $g(B_a,x_a) = \max \left[ 0, \, B_a - x_a \cdot w_a \right] - x_a \cdot u_a$, which is defined similarly to equation (\ref{g}), except that \emph{all} the variables incident on $a$ are considered instead of all but one; $g^*(B_a) = \min_{x_a} g(B_a, x_a)$; $\{x^*_a\}$ is the set of partial configurations $x_a$ that realize $g^*(B_a)$. When all the $\{u_{ka}\}$ are non-zero, this expression simplifies greatly:
\begin{eqnarray}
  E &=& \sum_a \max \left[ 0, \, B_a - x^*_a \cdot w_a \right] \label{E}
\end{eqnarray}
where $x^*_a$ is \emph{any} of the (possibly degenerate) partial configurations that minimize $g(B_a,x_a)$. Such a configuration, in turn, is easy to compute if all the $\{h_{ka}\}$ are different from zero, as can be seen by writing explicitly the minimization with respect to one of the variables, say $x_{ka}$:
\begin{eqnarray}
  \min_{x_a} g(B_a, x_a) &=&\min \biggl\{ \min_{x_{a \setminus k}} \left[ \max \left( 0, \, B_a - x_{a \setminus k} \cdot w_{a \setminus k} \right) - x_{a \setminus k} \cdot u_{a \setminus k} \right] , \, \nonumber \\
  && \min_{x_{a \setminus k}} \left[ \max \left( 0, \, B_a - w_{ka} - x_{a \setminus k} \cdot w_{a \setminus k} \right) - x_{a \setminus k} \cdot u_{a \setminus k} - u_{ka} \right] \biggr\} \\
  &=& \min_{x_{a \setminus k}} \left[ \max \left( 0, \, B_a - x_{a \setminus k} \cdot w_{a \setminus k} \right) - x_{a \setminus k} \cdot u_{a \setminus k} \right] + \min (0, -h_{ka}-u_{ka})
\end{eqnarray}
where the choice of the first term in the two-term minimum corresponds to $x^*_{ka} = 0$ while that of the second corresponds to $x^*_{ka} = 1$, provided $h_{ka} \neq 0$. Under the combined assumption $u_{ka} \neq 0 \neq h_{ka} \ \forall (ka)$ we have $x^*_{ka} = \theta(h_{ka} + u_{ka})$, where $\theta(x) = 1$ for $x>0$ and $\theta(x) = 0$ otherwise.

The expression of the entropy is somewhat more complicated:
\begin{eqnarray}
  S &=& \sum_a \left\{ \log \sum_{x_a \in \{x^*_a\}} \mathrm e^{x_a \cdot \eta_a} - \frac { \sum_{x_a \in \{x^*_a\}} x_a \cdot \eta_a \ \mathrm e^{x_a \cdot \eta_a} } { \sum_{x_a \in \{x^*_a\}} \mathrm e^{x_a \cdot \eta_a} } \right\} + \nonumber \\
  && + \sum_k \left\{ \log \sum_{a \in \partial k : h_{ka} = h^*_k} \mathrm e^{\xi_{ka}} - \frac {\sum_{a \in \partial k : h_{ka} = h^*_k} \xi_{ka} \mathrm e^{\xi_{ka}}} {\sum_{a \in \partial k : h_{ka} = h^*_k} \mathrm e^{\xi_{ka}}} \right\} +  \nonumber \\
  && - \sum_{ka} \mathbbm 1 \left[ h_{ka} + u_{ka} = 0 \right] \left\{ \log \left( 1+\mathrm e^{\xi_{ka} + \eta_{ka}} \right) - \frac {\left(\xi_{ka} + \eta_{ka}\right) \mathrm e^{\xi_{ka} + \eta_{ka}}} {1+\mathrm e^{\xi_{ka} + \eta_{ka}}} \right\} \label{av_entropy}
\end{eqnarray}
with the same notations as above and $h^*_k \equiv \max_{a \in \partial k} h_{ka}$.

Once the fixed-point fields $\{h_{ka},\xi_{ka}\}$ (or equivalently $\{u_{ka},\eta_{ka}\}$) are known, it is straightforward to compute the other set of fields via (\ref{u_update}, \ref{eta_update}) or via (\ref{h_update}, \ref{xi_update}), and then to compute the energy and entropy of the ground states via (\ref{av_energy}, \ref{av_entropy}).

\section{An efficient algorithm for random instances of the problem}

\subsection{Characterization of the random ensemble of instances} \label{random_ensemble}

An instance of the problem can be represented as a bipartite graph $\mathcal I = (K, A; E)$ in which $K = \{1, \cdots, N_\mathrm k\}$ is the set of keywords, $A = \{1,\cdots,N_\mathrm a\}$ is the set of advertisers, and $E = \{1,\cdots,N_\mathrm e\}$ is the set of the edges $(ka)$ such that $w_{ka} > 0$ (notice that this graph should not be confused with the factor graph introduced earlier). We assume the graph to be random with given $N_\mathrm a$, $N_\mathrm k$ and $N_\mathrm e$, and that in the thermodynamic limit $N_\mathrm a, N_\mathrm k, N_\mathrm e \to \infty$ the average connectivities $c_\mathrm a \equiv N_\mathrm e / N_\mathrm a$ and $c_\mathrm k \equiv N_\mathrm e / N_\mathrm k$ both remain finite. We consider the (non-zero) bids $w_{ka}$ to be iid random variables with uniform distribution in $]0,1]$. Given the bids $\{w_{a1},\cdots,w_{ac}\}$ offered by an advertiser $a$ with connectivity $c$, any budget smaller than $B_a^{\mathrm{min}} \equiv \min_j w_{aj}$ or larger than $B_a^{\mathrm{max}} \equiv \sum_j w_{aj}$ will be meaningless. We shall therefore consider a distribution of budgets \emph{correlated} to the connectivities and to the bids, in such a way that each budget $B_a$ is constrained to the interval $[B_a^\mathrm{min},B_a^\mathrm{max}]$. The most ``natural'' distribution of budgets is obtained by conditioning the \emph{uniform} distribution in $\bigotimes_a[B_a^\mathrm{min},B_a^\mathrm{max}]$ to have a given sum, specified by a parameter $\bar b \in [0,1]$:
\begin{eqnarray}
  \mathbbm P \left. \left[ \{B_a\} \right| \bar b, \{w_{ka}\} \right] \propto \delta \left( \sum_a \left[ B_a^\mathrm{min} + \bar b \left( B_a^\mathrm{max} - B_a^\mathrm{min} \right) \right] - \sum_a B_a \right) \prod_a \theta(B_a - B_a^\mathrm{min}) \, \theta(B_a^\mathrm{max} - B_a) \, .
\end{eqnarray}
As $N_\mathrm a \rightarrow \infty$ the ``reduced'' budgets $b_a \equiv (B_a - B_a^\mathrm{min}) / (B_a^\mathrm{max} - B_a^\mathrm{min})$ become iid random variables with an exponential distribution in the interval $[0,1]$ with parameter $\gamma$ given by the (unique) solution of $1 + 1/(\mathrm e^\gamma - 1) - 1/\gamma = \bar b$. This makes it very simple to generate large instances from this distribution. In the thermodynamic limit the average budget $\bar B$ is related to the parameter $\bar b$ by the linear relation:
\begin{equation} \label{B_vs_b}
  \bar B = \frac{\mathrm e^{-c_\mathrm a}} {c_\mathrm a} \left( \mathrm e^{c_\mathrm a} - 1 - c_\mathrm a \right) \left(1 - \bar b\right) + \frac{c_\mathrm a}{2} \bar b \, .
\end{equation}

We shall denote this random ensemble as $\mathscr E(N_\mathrm a, N_\mathrm k, N_\mathrm e, \bar B)$ for finite-size instances and as $\mathscr E_\infty(c_\mathrm a, c_\mathrm k, \bar b)$ in the thermodynamic limit.
In the following sections, we shall discuss how the properties of $\mathscr E$ and $\mathscr E_\infty$ vary with these parameters. Of course, one can consider many other ensembles of instances. For example, one could consider an exponential distribution for the bids, or remove the correlation between bids and budgets. In fact, only the distribution of the \emph{actual} bids and budgets can claim to be more interesting than the other possible ones, and the only motivation for our choice is that it is reasonable and simple.

\subsection{Definition of the algorithm} \label{algorithm}

We shall now define an algorithm, based on the analysis of the previous section, which finds efficiently optimal (or almost-optimal) configurations for random instances of the problem extracted from the distribution described in the previous paragraph, and which coincides with the zero-temperature Belief Propagation (BP) algorithm, with the cavity fields playing the role of messages, and the self-consistent equations for the cavity fields playing the role of update rules. We shall see that in order to find an optimal configuration of the problem it is sufficient to compute the hard fields, so the update equations for the messages will be (\ref{u_update}) and (\ref{h_update}), which we rewrite as $h_{ka} = \mathcal H_{ka}(u_{a \setminus k})$ and $u_{ka} = \mathcal U_{ka}(h_{k \setminus a})$. For the sake of concreteness the pseudo-code of the Belief Propagation algorithm is written in Algorithm \ref{algo:BP}. The inputs are an off-line AdWords instance $\mathcal I$, an initial set of messages $\{h_{ka}^0\}$, a convergence criterion $\mathscr C(\{h_{ka}^t\},\epsilon) \in \{\texttt{true}, \texttt{false}\}$, and a maximum number of iterations $T$.

\begin{algorithm}
\begin{pseudocode}[shadowbox]{Zero-temperature BP}{\mathcal I, \{h_{ka}^0\}, \mathscr C,T}
\label{algo:BP}
t \GETS 0\\
\WHILE \big(\neg \mathscr C(\{h_{ka}^t\},\epsilon)\big) \AND \big(t < T\big) \DO
\BEGIN 
  \FOREACH (ka) \in E \DO
  u^t_{ka} \GETS \mathcal U_{ka}(h_{k \setminus a}^t)\\
  t \GETS t+1 \\
  \FOREACH (ka) \in E \DO
    h_{ka}^t \GETS \mathcal H_{ka}(u_{a \setminus k}^{t-1})\\
\END\\
\FOREACH (ka) \in E \DO
  u_{ka}^t \GETS \mathcal U_{ka}(h_{k \setminus a}^t) \\
\IF \big(\mathscr C(\{h_{ka}^t\},\epsilon) \big) \AND \big(h_{ka}^t + u_{ka}^t \neq 0 \ \ \forall (ka) \big) \THEN
\BEGIN
  \FOREACH ka\in E \DO 
    x_{ka} \GETS \theta(h_{ka}^t + u_{ka}^t)\\
  \RETURN {x}
\END
\ELSE
\RETURN {\mbox{Undetermined}}
\end{pseudocode}
\end{algorithm}

This algorithm is successful in finding an optimal configuration only if the following two conditions are met: the messages must converge to a fixed point, and at the fixed point $h_{ka} + u_{ka}$ must be non-zero for all $(ka)$. As we shall see in the following section, these conditions are not always verified for random instances extracted from the random ensemble $\mathscr E(N_\mathrm a, N_\mathrm k, N_\mathrm e, \bar B)$ described in paragraph \ref{random_ensemble}. However, we shall see that it is possible to modify the update equations (\ref{u_update}) and (\ref{h_update}) so that a solution is (almost) always found.

\section{Numerical results}

\subsection{Convergence of the zero-temperature BP} \label{convergence}

We have implemented the Algorithm \ref{algo:BP} and used it to solve random instances from the distribution $\mathscr E$ described in paragraph \ref{random_ensemble}, with several values for the distribution parameters $N_{\mathrm a}$, $N_{\mathrm k}$, $N_\mathrm e$ and $\bar b$.

\begin{figure}
  \input{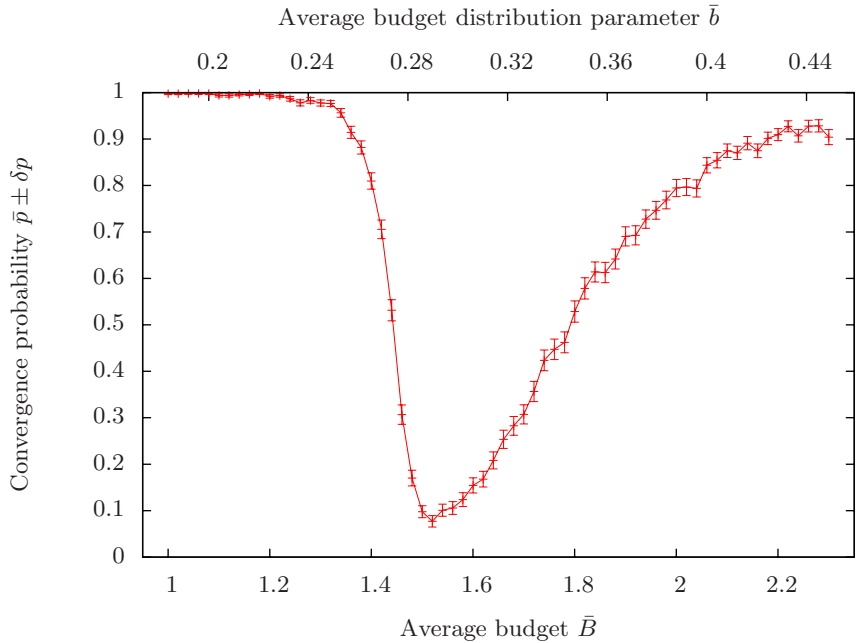}
  \caption{Convergence of the algorithm for different values of the average budget $\bar B$. The other parameters of the instance distribution are as follows: $N_\mathrm a = 1\,000$, $N_\mathrm k = 3\,000$ and $N_\mathrm e = 10\,000$ (the parameters for the algorithm are specified in the text). The samples are grouped according to their average budget $\bar B$ in bins of amplitude $0.02$ (the horizontal errorbars are not shown for simplicity); the corresponding values of the distribution parameter $\bar b$, computed according to (\ref{B_vs_b}), are shown on the upper scale. The number of samples in each bin is approximately 500.}
 \label{convergence_plot}
\end{figure}

\begin{figure}
  \input{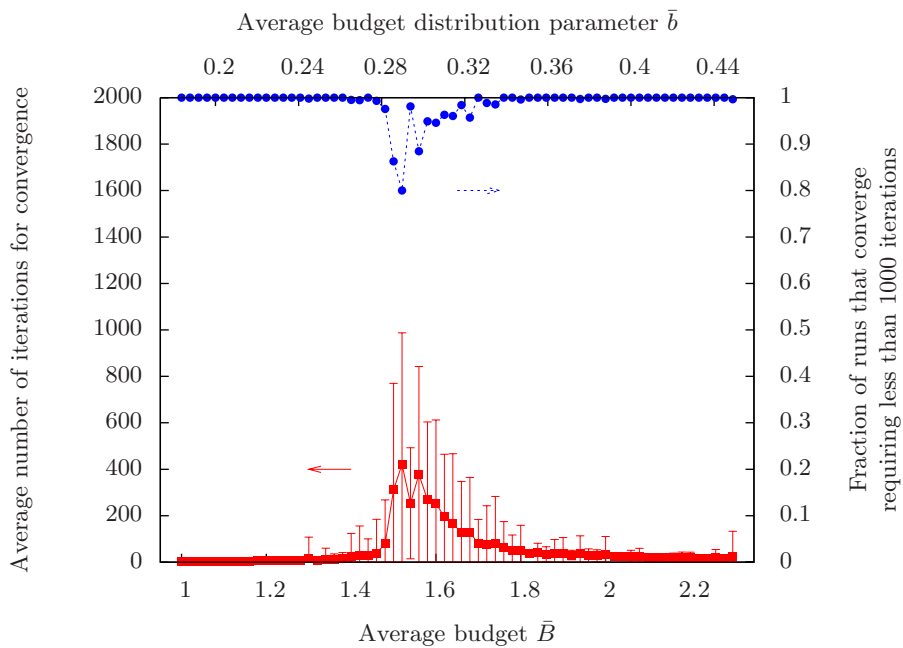}
  \caption{Number of iterations required for convergence (same samples as in Figure \ref{convergence_plot}). The bottom curve (left scale) shows the average number of iterations required for convergence and its standard error. The upper curve (right scale) shows the fraction of the samples that converge requiring less than 1\,000 iterations.}
 \label{iterations_plot}
\end{figure}

In Figures \ref{convergence_plot} and \ref{iterations_plot} we study the convergence behavior of the algorithm as the budget distribution parameter $\bar b$ is varied, while the other parameters are kept fixed: $N_\mathrm a = 1\,000$, $N_\mathrm k = 3\,000$, $N_\mathrm e = 10\,000$. The algorithm is run with the following parameters: the initial values of the messages are set to $h_{ka}^0 = 0$ for all $(ka)$, the maximum number of iteration is $T = 2\,000$, the convergence criterion is
\begin{equation}
  \label{convergence_criterion}
  \mathscr C (\{h_{ka}^t\}, \epsilon) = \Biggl \{ \sum_{(ka)} \left[ \left| h_{ka}^t - h_{ka}^{t-1} \right|^2 + \left| u_{ka}^t - u_{ka}^{t-1} \right|^2 \right] < \epsilon^2 \Biggr \} \wedge \Biggl \{ \theta(h_{ka}^t + u_{ka}^t) = \theta(h_{ka}^{t-1} + u_{ka}^{t-1}), \ \forall (ka) \Biggr \}
\end{equation}
with $\epsilon = 10^{-5}$. A total of 31\,736 instances are generated with values of $\bar b$ in the range $[0.17,0.45]$, and they are grouped according to the \emph{actual} value of the average budget $\bar B$, in bins of amplitude 0.02 (recall that since the $N_\mathrm a$ individual budgets are extracted as iid random variables from a distribution with parameter $\bar b$, their average $\bar B$ is also a random variable).

For a given bin, if the total number of samples is $n$ and the number of samples for which convergence is obtained is $m$, the distribution for the probability (over the choice of the instance) of convergence $p$ is computed as follows\footnote{Assuming a uniform prior over $p$.}:
\begin{eqnarray}
  \mathbbm P [m | n,p] = {n \choose m} p^m (1-p)^{n-m} \Rightarrow
  \mathbbm P [p | n,m] = (n+1) {n \choose m} p^m (1-p)^{n-m}
\end{eqnarray}
so that the expected value of the convergence probability $\bar p$ and its standard deviation $\delta p$ are:
\begin{eqnarray}
  \bar p &=& \frac {m+1}{n+2} \, , \\
  \delta p &=& \sqrt{\frac{m+1}{n+2} \left(\frac{m+2}{n+3} - \frac{m+1}{n+2} \right)} \, .
\end{eqnarray}

\begin{figure}
  \input{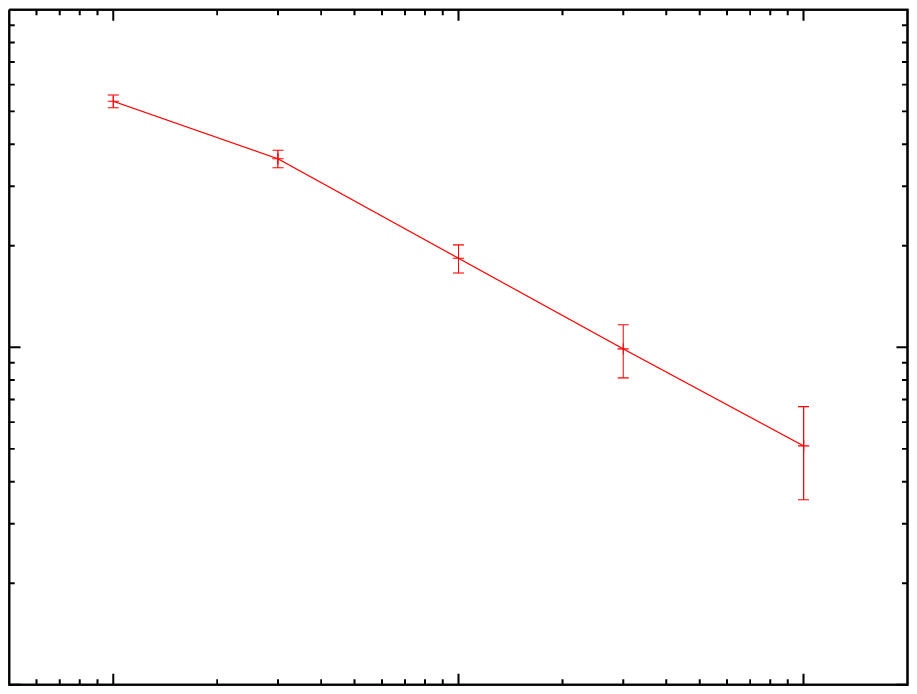}
  \caption{Scaling of the convergence probability $\bar p$ with the number of advertisers $N_\mathrm a$ for fixed $N_\mathrm k / N_\mathrm a = 3$, $N_\mathrm e / N_\mathrm a = 10$ and $\bar B = 1.44$. The data points are very well fitted by $\bar p \propto N_\mathrm a ^ \nu$ with the exponent $\nu = -1/2$, indicating a sharp transition in the infinite size limit.}
  \label{scaling_plot}
\end{figure}

In Figure \ref{convergence_plot} we show the results for $\bar p$ and $\delta p$ as a function of $\bar B$. For small values, $\bar B \lesssim 1.4$, the algorithm almost always converges, while for $1.4 \lesssim \bar B \lesssim 1.5$ there is a sharp decrease in the probability of convergence. For larger values, $1.5 \lesssim \bar B$, the probability of convergence gradually increases back to 1, reaching the value $\bar p \simeq 0.5$ for $\bar B \simeq 1.8$.

The equivalence of the convergence of our algorithm with the RS assumption together with the above results suggest a phase diagram (in terms of $\bar B$) with a RS phase for $\bar B < \bar B_0$, a Replica Symmetry Broken (RSB) phase for $\bar B_0 < \bar B < \bar B_1$ and again a RS phase for $\bar B_1 < \bar B$, where $\bar B_0$ and $\bar B_1$ are functions of the average connectivities $c_\mathrm a$ and $c_\mathrm k$ and are roughly equal, for $c_\mathrm a = 10$ and $c_\mathrm k = 10/3$, to 1.4 and 1.8 respectively. The existence of a phase transition at $\bar B \simeq 1.4$ is supported by the scaling of $\bar p$ with the size of the instance (i.e. $N_\mathrm a$) shown in Figure \ref{scaling_plot}, which indicates that slightly above $\bar B = 1.4$ in the infinite size limit $\bar p$ tends to zero. The transition at $\bar B = 1.8$ on the other hand does \emph{not} appear to be sharp. This is not in contradiction with the suggested phase diagram, since the convergence probability on finite size instances might be affected by the presence of local structures where the fields do not reach convergence, but which do not contribute to the properties of the ensemble in the infinite size limit, as happens for example in the random $(2+p)$-XORSAT.
Indeed, we shall see that the analysis of the stability of the RS solution (in paragraph \ref{stability}) confirms this phase diagram.

It should be stressed that these results on convergence could, in principle, depend heavily on the choices of the convergence criterion $\mathscr C(\{h_{ka}^t\},\epsilon)$, of the maximum number of iterations allowed $T$, and of the initial values of the messages $h_{ka}^0$.
We find that the choice of the convergence criterion has very limited impact, as long as it is a reasonable one. We experimented with several values of $\epsilon$ between $10^{-10}$ and $10^{-1}$, and we also tried other possible criteria, such as the following two:
\begin{eqnarray}
  \mathscr C' (\{h_{ka}^t\},\epsilon) &=& \Biggl \{ \max_{(ka)} \left[ \max \left( \left| h_{ka}^t - h_{ka}^{t-1} \right|, \left| u_{ka}^t - u_{ka}^{t-1} \right| \right) \right] < \epsilon \Biggr \} \wedge \Biggl \{ \theta(h_{ka}^t + u_{ka}^t) = \theta(h_{ka}^{t-1} + u_{ka}^{t-1}), \ \forall (ka) \Biggr \}  \, , \\
  \mathscr C''(\{h_{ka}^t\},\epsilon) &=& \biggl\{ \theta(h_{ka}^t + u_{ka}^t) = \theta(h_{ka}^{t-1} + u_{ka}^{t-1}) = \cdots = \theta(h_{ka}^{t-4} + u_{ka}^{t-4}), \ \forall (ka) \biggr\} \, . \label{criterion_c''}
\end{eqnarray}
In all these cases, we find very similar results.
We also tried an alternative for the initial condition: setting $h_{ka}^0$ to random values extracted uniformly in the interval $[0,1[$. Again, this has very limited impact on the convergence behavior of the algorithm.
Finally, we analyzed the possible effect of the cut-off on the number of iterations allowed. In Figure \ref{iterations_plot} we show the average number of iterations $n_i$ required for convergence, together with the fraction of runs that converge and require less than 1\,000 iterations (i.e. half the maximum number of iterations allowed, $T$). By comparing the two plots, and assuming that the distribution of $n_i$ is unimodal when $T \to \infty$, it seems clear that the effect of this cut-off is negligible for the samples considered.

\begin{figure}
  \input{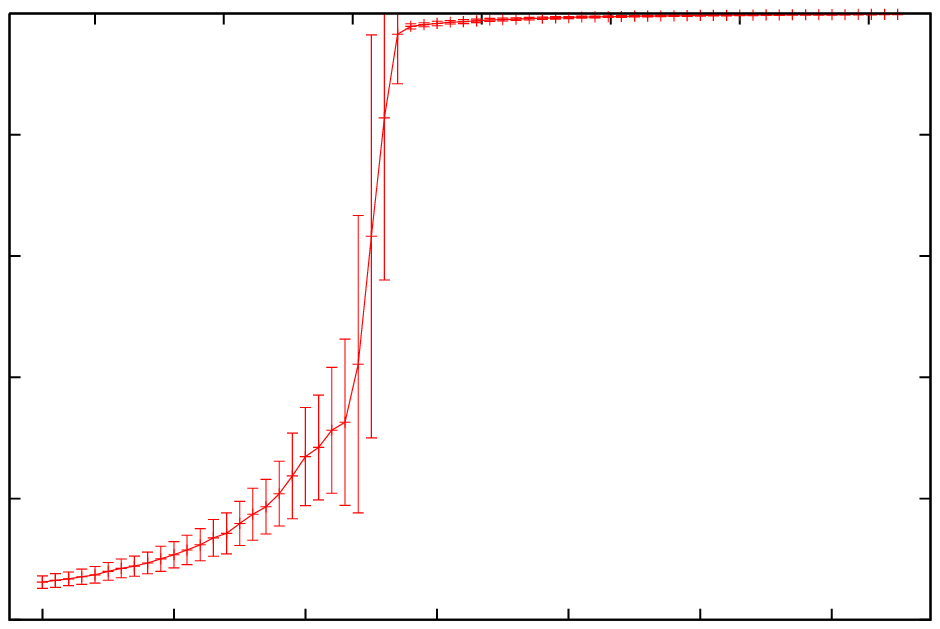}
  \caption{Fraction of \emph{frozen} variables, i.e. $x_{ka}$ such that $h_{ka} + u_{ka} \neq 0$ (same samples as Figures \ref{convergence_plot} and \ref{iterations_plot}).}
  \label{frozen_plot}
\end{figure}

In Figure \ref{frozen_plot}, we see that for $\bar B \lesssim 1.5$ a large fraction of the variables are unfrozen, that is to say that the fields associated to them verify $h_{ka} + u_{ka} = 0$, so that the simple prescription used by Algorithm \ref{algo:BP} is unable to assign them.
The existence of regions of values of $\bar B$ where the messages often do not converge to a fixed point, or where the fixed point corresponds to a large number of unfrozen variables, constitutes a serious problem for the practical use of our algorithm. However, as we mentioned before, it is possible to ``fix'' it with two simple modifications of the update equations (\ref{u_update}) and (\ref{h_update}).

The probability of convergence $\bar p$ of the algorithm is greatly improved if the update rule for the fields $h^t_{ka}$ is modified to include a \emph{reinforcement} term \cite{Braunstein05}. The modified rule takes the form:
\begin{equation} \label{reinforcement}
  h^{t}_{ka} \leftarrow \mathcal H_{ka}(u_{a \setminus k}^{t-1}) + \rho (h^{t-1}_{ka} + u^{t-1}_{ka})
\end{equation}
where $\rho$ is a constant real number which we shall refer to as \emph{reinforcement parameter}. In practice, the value taken by the new field is a linear combination of the value prescribed by the update equation (\ref{h_update}) and the value of the \emph{full} field (as opposed to the cavity one) acting on the variable being updated, which is equal to $h^{t-1}_{ka} + u^{t-1}_{ka}$. The coefficient $\rho$ is used to ``tune'' the strength of this term.

\begin{figure}
  \input{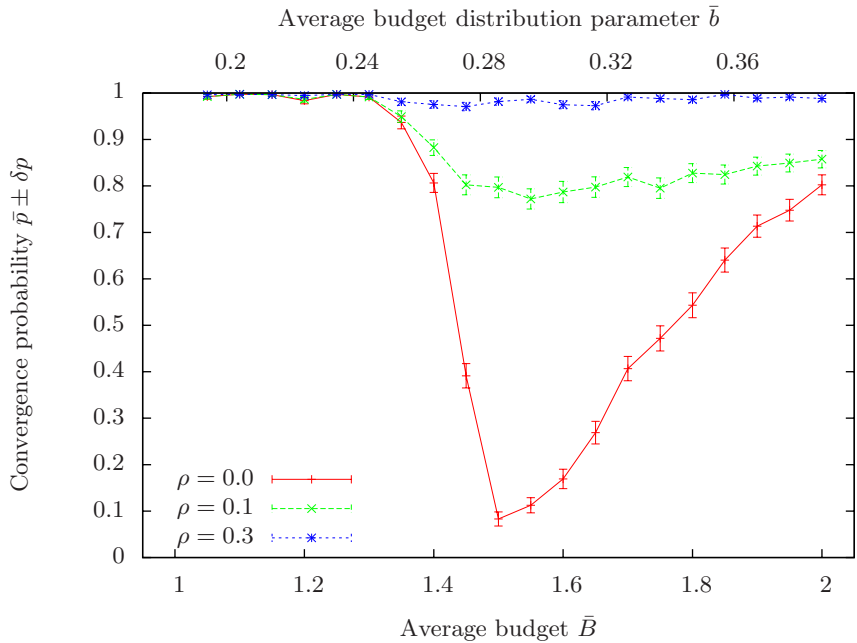}
  \caption{Effect of reinforcement on the convergence of the algorithm. The parameters of the instances distribution are the same as in Figure \ref{convergence_plot}, but a reinforcement term with parameter $\rho$ is included in the update equation (\ref{reinforcement}).}
  \label{reinforcement_plot}
\end{figure}

The results of this modification in the update rule are quite spectacular, as can be seen on Figure \ref{reinforcement_plot}, where the probability of convergence $\bar p$ is compared, for the \emph{same} samples, with different values of the reinforcement parameter $\rho$. The parameters of the distribution of instances are the same as in Figures \ref{convergence_plot} -- \ref{frozen_plot}, i.e. $N_\mathrm a = 1\,000$, $N_\mathrm k = 3\,000$, $N_\mathrm e = 10\,000$, and the total number of instances is 6\,771. The parameters for the algorithm are as follows: the convergence criterion used is $\mathscr C''$, defined in (\ref{criterion_c''}), the maximum number of iterations allowed is $T=800$, and the initial values of the messages are $h_{ka}^0 = 0$ for all $(ka)$. For $\rho = 0.3$, approximately 99\% of the samples converge within 500 iteration.

\begin{figure}
  \input{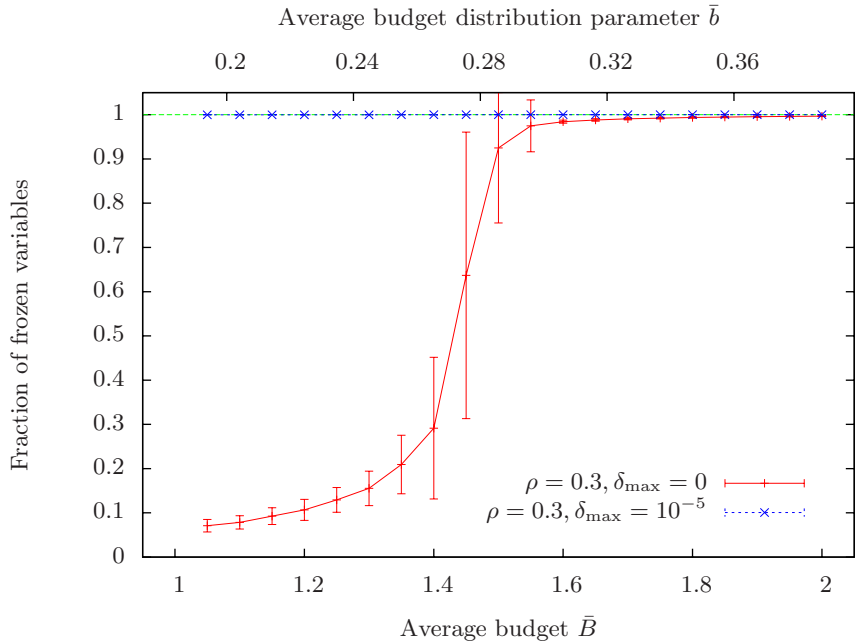}
  \caption{Effect of random pinning fields on the numbers of frozen variables. The instances are the same as in Figure \ref{reinforcement_plot}.}
  \label{noise_plot}
\end{figure}

The second problem of the original algorithm, namely that a large fraction of the variables are unfrozen for small values of $\bar B$, can also be easily solved, by adding a random \emph{pinning field} to the update equations, which become (including the reinforcement term):
\begin{eqnarray}
  h^{t}_{ka} &\leftarrow& \mathcal H_{ka}(u_{a \setminus k}^{t-1}) + \rho (h^{t-1}_{ka} + u^{t-1}_{ka} - \delta_{ka}) + \delta_{ka} \label{h_rein_noise} \, , \\
  u^{t}_{ka} &\leftarrow& \mathcal U_{ka}(h_{k \setminus a}^{t}) + \delta_{ka} \label{u_rein_noise}
\end{eqnarray}
where the pinning fields $\delta_{ka}$ are constant random real numbers, extracted from a uniform distribution in the interval $]0,\delta_{\max}]$.
Notice that the pinning field also enters in the reinforcement term: since it affects both the updates of $h_{ka}$ and $u_{ka}$, its effect is duplicated in the sum $h_{ka} + u_{ka}$, which must therefore be corrected as $h_{ka} + u_{ka} - \delta_{ka}$ (this also applies to the step functions defining the values of $x_{ka}$ and the convergence criteria). Figure \ref{noise_plot} shows the effect of the pinning field on the average number of frozen variables: all the variables become frozen, except at most 6 out of 10\,000 in the samples considered.

With these two modifications (i.e. the introduction of a reinforcement term and of random pinning fields), our algorithm is almost always capable to find a low-energy configuration of the problem. However, the introduction of these ``spurious'' terms in the update equations could lead to an increase in the energy of the configuration obtained, relative to the true optimal energy of the instance. In order to verify that this effect is not significant, we have compared the energy of the configuration found by the algorithm using both reinforcement and pinning fields, with the average energy computed according to equation (\ref{av_energy}) in the absence of them, for those instances for which convergence is obtained. Notice that this requires to compute the \emph{soft} fields $\eta_{ka}$ and $\xi_{ka}$ in addition to the hard ones, which we have done for the sake of the comparison.

\begin{figure}
  \input{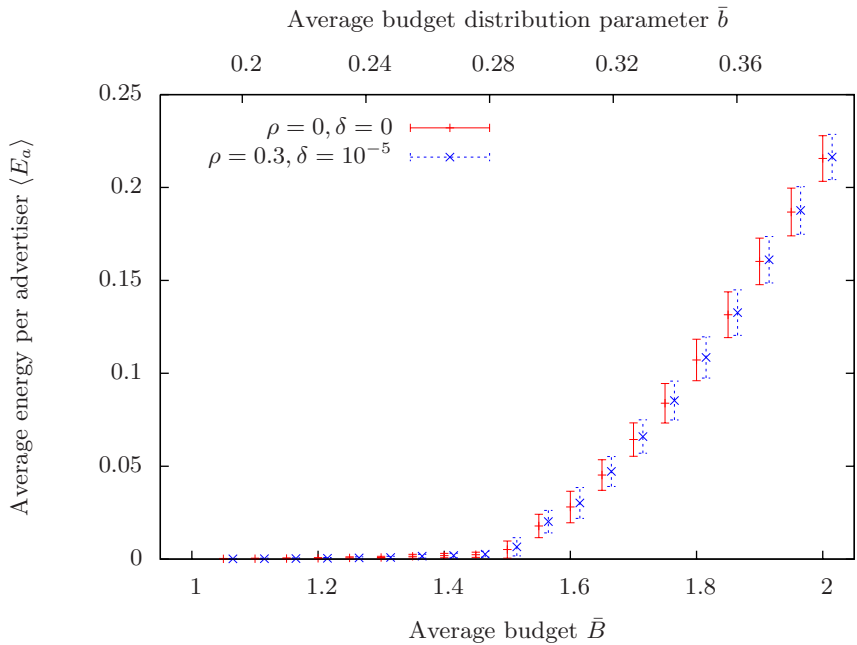}
  \caption{Comparison of the average energy per advertiser obtained with reinforcement and pinning fields to that obtained without them. The instances are the subset of those analyzed in Figure \ref{reinforcement_plot} for which the algorithm without reinforcement and pinning field converges; their total number is 4\,914. The points corresponding to $\rho=0.3, \delta_{\max} = 10^{-5}$ have been slightly shifted horizontally to make the comparison possible.}
  \label{energy_plot}
\end{figure}

In Figure \ref{energy_plot} we show the results of this comparison: the addition of reinforcement and pinning fields causes a systematic but very small increase in the average energy per advertiser. In order to quantify the relevance of this effect, its magnitude should be compared to the average \emph{revenue} per advertiser rather than to the value of the energy. We find that the relative effect on the revenue is always smaller than $3 \times 10^{-3}$, with an average value and a standard deviation of $3 \times 10^{-4} \pm 4 \times 10^{-4}$.

\subsection{Stability of the replica symmetric solution and phase diagram of the problem} \label{stability}

As we have seen in the discussion of paragraph \ref{convergence}, there is an interval of values of the parameter of the budget distribution $\bar B$ for which the algorithm (in the absence of reinforcement) does not converge, and this suggests the presence of a replica symmetry broken phase. In this paragraph, we are going to present a method which permits to verify this hypothesis directly, and to present some numerical results confirming it. This method was originally introduced in \cite{Montanari03} to verify the stability of 1-step RSB (1RSB) solutions, but it is easily simplified to verify the stability of RS solutions as well, as was done for instance in \cite{Marinari06, Raymond07}.

The idea behind the method is very simple. The RS solution is a special case of a more general solution, the 1RSB solution. The difference between the two is the following: while in the RS solution the value of the cavity fields $h_{ka}$ at the fixed point are well defined, in the 1RSB solution they are defined only \emph{in distribution}. The correct parametrization of the 1RSB solution is not given by a set of \emph{numbers} $\{
h_{ka}\}$, but by a set of \emph{functions} $\{\mathcal P_{ka}(h_{ka})\}$ which describe the probability distribution of the values of $h_{ka}$. The RS solution is a special case of this, in which all the functions $\mathcal P_{ka}(\cdot)$ are delta functions.

A necessary condition for the RS solution to be correct is that it is \emph{stable} under the enlargement of the solution space to the 1RSB case. This can be verified by considering what happens under the iteration (\ref{u_update}, \ref{h_update}) if instead of a sharply defined value the cavity fields $h_{ka}$ have some small but finite variance $v_{ka}/2$ around their fixed point values $h^*_{ka}$, i.e. if they have a distribution
\begin{equation}
  \mathcal P_{ka}(h_{ka}) = \mathcal N_{ka} \, \mathrm e^{-(h_{ka} - h^*_{ka})^2/v_{ka}}
\end{equation}
where $\mathcal N_{ka}$ is a normalization. Under the iteration (\ref{u_update},\ref{h_update}), the field $h_{jb}$ is computed as a function $\mathscr H_{jb}$ of the fields $h_{ka}$ that are at distance 2 from $h_{jb}$ on the graph representing the instance, i.e.
\begin{equation}
  \mathscr H_{jb}(\{h_{ka}\}) = \mathcal H_{jb} \left( \{\mathcal U _{kb}(h_{k \setminus b}) | k \in \partial b \setminus j \} \right) \; .
\end{equation}

If $\{h_{ka}\}$ are close to the fixed point values $\{h^*_{ka}\}$, we can expand
\begin{equation}
  h_{jb} = \mathscr H_{jb} \left( \{h^*_{ka}\} \right) + \sum_{k \in \partial b \setminus j} \sum_{a \in \partial k \setminus b} \frac { \partial \mathscr H_{jb} \left( \{h^*_{ka}\} \right) } {\partial h_{ka}} (h_{ka} - h^*_{ka}) + \cdots
\end{equation}
where $\mathscr H_{jb} \left( \{h^*_{ka}\} \right) = h^*_{jb}$. The probability distribution of the values of $h_{jb}$ is then given by
\begin{equation}
 \mathcal P_{jb}(h_{jb}) = \mathbbm P \left[ h^*_{jb} + \sum_{k \in \partial b \setminus j} \sum_{a \in \partial k \setminus b} \frac { \partial \mathscr H_{jb} \left( \{h^*_{ka}\} \right) } {\partial h_{ka}} (h_{ka} - h^*_{ka}) = h_{jb} \right]
\end{equation}
which can be computed by writing the integral representation of the delta function representing this probability and performing twice a gaussian integration. This gives
\begin{equation}
 \mathcal P_{jb}(h_{jb}) = \mathcal N_{jb} \, \mathrm e^{-(h_{jb} - h^*_{jb})^2/v_{jb}}
\end{equation}
where we identify
\begin{equation} \label{v_update}
  v_{jb} = \sum_{k \in \partial b \setminus j} \sum_{a \in \partial k \setminus b} \left( \frac { \partial \mathscr H_{jb} \left( \{h^*_{ka}\} \right) } {\partial h_{ka}} \right)^2 v_{ka} \; .
\end{equation}

In order to verify the stability of the RS solution towards 1-step replica symmetry breaking, one can then associate a real number $v_{ka}$ to each cavity field $h_{ka}$, representing the variance of its distribution, and update the values of $v_{ka}$ with equation (\ref{v_update}) each time that the field $h_{ka}$ is updated. As the cavity fields reach their fixed point values $\{h^*_{ka}\}$, one can perform one more iteration for all the edges $\{(ka)\}$ in the graph and compare the values of $\{v_{ka}\}$ before the iteration with those $\{v'_{ka}\}$ after it, and measure the \emph{stability parameter} $\lambda$:
\begin{equation} \label{lambda}
  \lambda \equiv \frac {\sum_{k,a} v'_{ka}} {\sum_{k,a} v_{ka}} \;.
\end{equation}
If $\lambda > 1$ the RS solution will be unstable towards a 1RSB enlargement of the solution space, otherwise it will be stable.

Notice that, in general, the derivative $\partial \mathscr H_{jb} / \partial h_{ka}$ will involve also the \emph{soft} field $\{\xi_{ka}\}$, which therefore must also be computed. This can be seen by writing explicitly
\begin{eqnarray}
  \left( \frac{\partial \mathscr H_{jb}}{\partial h_{ka}} \right)^2 &=&
    \sum_{k \in \partial b \setminus j} \left( \frac{d \mathcal H_{jb}}{d u_{kb}} \right)^2 \sum_{a \in \partial k \setminus b} \left( \frac{d \mathcal U_{kb}}{d h_{ka}} \right)^2  \\
  &=& \sum_{k \in \partial b \setminus j} \left\{ \frac{ \sum_{ x_{b \setminus j} \in x^*_{b \setminus j}(B_b)} x_{kb} \; \mathrm e^{ x_{b \setminus j} \cdot \eta_{b \setminus j}} } { \sum_{ x_{b \setminus j} \in x^*_{b \setminus j}(B_b)}  \mathrm e^{ x_{b \setminus j} \cdot \eta_{b \setminus j}} } - \frac{ \sum_{ x_{b \setminus j} \in x^*_{b \setminus j}(B_b-w_{jb})} x_{kb} \; \mathrm e^{ x_{b \setminus j} \cdot \eta_{b \setminus j}} } { \sum_{ x_{b \setminus j} \in x^*_{b \setminus j}(B_b-w_{jb})}  \mathrm e^{ x_{b \setminus j} \cdot \eta_{b \setminus j}} } \right\}^2 \times \nonumber \\
    && \times \sum_{a \in \partial k \setminus b} \mathbbm 1 \left[ h_{ka} = \max_{c \in \partial k \setminus b} h_{kc} \right]
\end{eqnarray}
where the soft fields $\eta_{b \setminus j}$ are computed as functions of $\xi_{k \setminus b}$ from (\ref{eta_update}), and where $x^*_{b \setminus j}(z)$ is the set of partial configurations that minimize the function $g(z,x_{b \setminus j})$ defined in (\ref{g}).

This procedure can be applied to any given instance. Alternatively, it can be used to verify if the RS solution is stable in the thermodynamic limit for the ensemble $\mathscr E_\infty(c_\mathrm a, c_\mathrm k, \bar b)$, using the \emph{population dynamics} technique: in this case, the topology of the problem is not fixed, but is extracted at random for each update. One then considers a population (i.e. a set) of $N$ triples $(h_i, \xi_i, v_i)$. At each step, a local tree of depth 2 is extracted, with average connectivity $c_\mathrm a$ for the root and $c_\mathrm k$ for the nodes at depth 1. Each leaf of the tree is associated at random with a triple $(h_i, \xi_i, v_i)$ from the population. The value of the fields $h_0$ and $\xi_0$ at the root are computed according to (\ref{u_update}--\ref{xi_update}), and the value of $v_0$ is computed according to (\ref{v_update}). Then, a triple $(h_i, \xi_i, v_i)$ is selected at random from the population and is replaced with $(h_0, \xi_0, v_0)$. This procedure is repeated a number of times $T_0 \equiv t_0 N \gg N$ (the population size) in order to reach the equilibrium distribution of fields $\mathcal P(h)$. The population of variances is then reinitialized to $v_{ka} = 1 \ \forall (ka)$, and another $T_1 \equiv t_1 N \gg N$ iterations are performed (renormalizing the variances after each $\Delta T = N$ iterations) so that the distribution of variances also equilibrates. Finally, $T_2 \equiv t_2 N \gg N$ iterations are performed, computing $\lambda_T$ according to (\ref{lambda}), and averaging the results.
Another quantity which is of interest to characterize the stability of the RS solution is the fraction of non-zero variances $v_{ka}$ in the population, which we shall denote by $\phi$. It can be computed together with $\lambda$, and it must be zero for the RS solution to be stable.

\begin{figure}
  \input{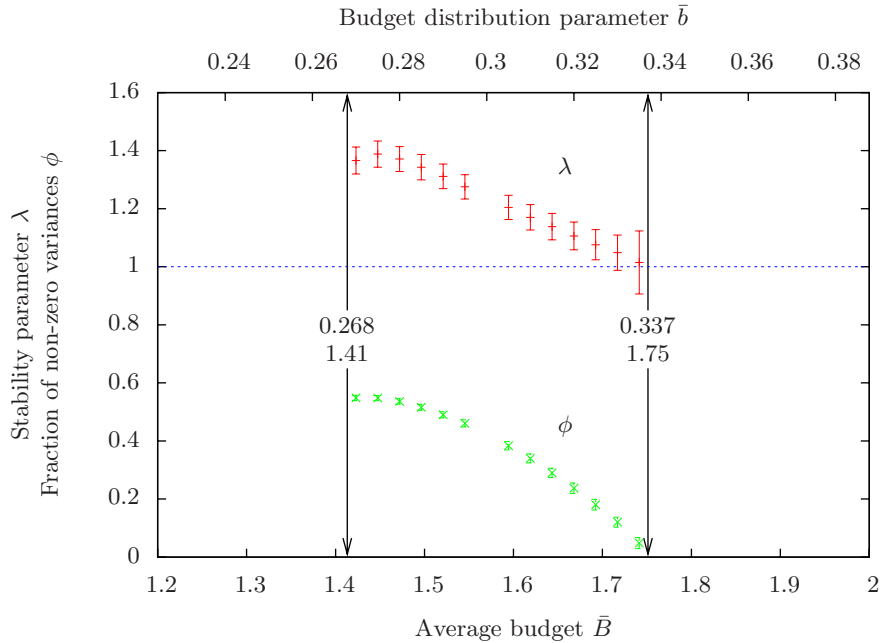}
  \caption{Stability parameter $\lambda$ and fraction of non-zero variances $\phi$ for different values of the budget distribution parameter $\bar b$, obtained from the population dynamics technique. The average connectivities used are the same as in the previous plots, i.e. $c_\mathrm a = 10$ and $c_\mathrm k = 10/3$. The region $0.268 \lesssim \bar b \lesssim 0.337$ (corresponding to $1.41 \lesssim \bar B \lesssim 1.75$) is unstable, i.e. has $\lambda > 1$. Outside of this region, both $\lambda$ and $\phi$ vanish. The values of $\lambda$ in the unstable region are well fitted by a line (except the first point), while the values of $\phi$ are well fitted by a parabola.}
  \label{lambda_plot}
\end{figure}

\begin{figure}
  \input{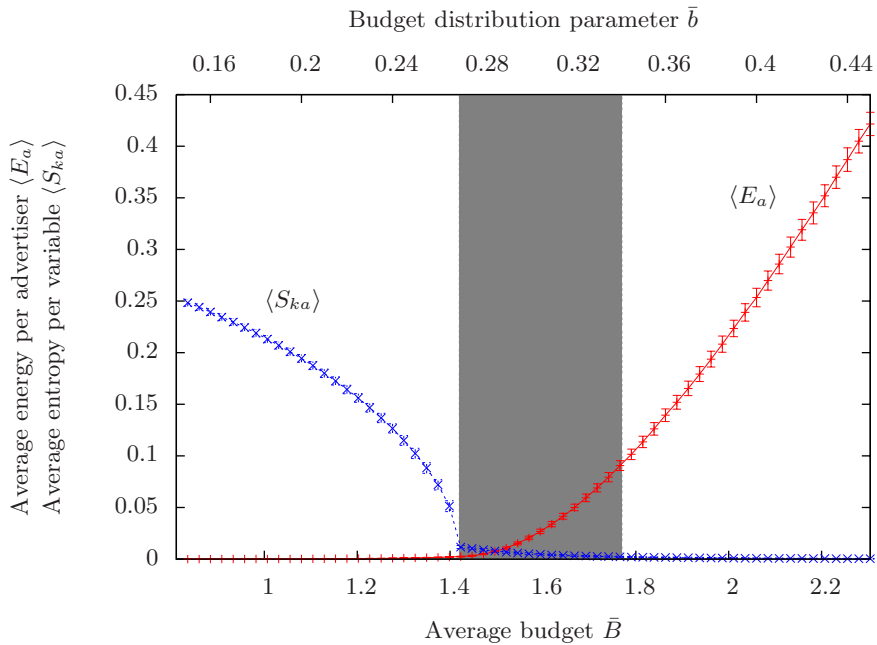}
  \caption{Average energy per advertiser and average entropy per variable computed with the population dynamics method. The average connectivities used are the same as in the previous plots, i.e. $c_\mathrm a = 10$ and $c_\mathrm k = 10/3$. The shaded region $\bar b \in [0.268, 0.337]$ corresponds to the RSB phase: in this region, the values obtained for $\left< E_a \right>$ and $\left< S_{ka} \right>$ in the RS approximation are (probably) not correct.}
  \label{energy_entropy_plot}
\end{figure}

Using this approach, we have obtained the results shown in Figure \ref{lambda_plot}. The parameters of the distribution of instances are the same as for the data shown in Figures \ref{convergence_plot}: average connectivity $c_\mathrm a = 10$ for the advertisers and $c_\mathrm k = 10/3$ for the keywords; the population size is $10^4$; the equilibration time is $t_0 = 1500$ for the distribution of fields and $t_1 = 1000$ for the distribution of variances; the measurement time is $t_2 = 4000$. We see that the region $\bar b \in [0.268,0.337]$ (corresponding to $\bar B \in [1.41,1.75]$) is unstable, i.e. has $\lambda > 1$. This is in very good agreement with the convergence results of the algorithm, which show a marked drop in the fraction of instances that converge in the same region. Figure \ref{energy_entropy_plot} shows the corresponding average energy per advertiser and the average entropy per variable. The nature of the two transitions appears different, and it may be interesting to study with more depth how ``sharp'' is the one at $B=1.75$.

The previous analysis confirms the phase diagram suggested in paragraph \ref{convergence}: for $\bar b < \bar b_0$ we have a RS phase with negligible average energy and positive entropy, for $\bar b_0 < \bar b < \bar b_1$ we have a RSB phase, and for $\bar b_1 < \bar b$ we have again a RS phase, but with positive energy and negligible entropy. The threshold values $\bar b_0$ and $\bar b_1$ depend on the average connectivities $c_\mathrm a$ and $c_\mathrm k$.

\subsection{Comparison to other algorithms}

We have compared zero-temperature BP to several other algorithms in terms of the average energy of the solutions they find. The alternative algorithms considered are Simulated Annealing (SA) and Linear Programming (LP) relaxation, with several rounding procedures. In the following paragraphs we shall briefly describe them, and then show and discuss the numerical results we obtained.

\subsubsection{Simulated annealing} 

Simulated annealing is an optimization technique first proposed by Kirkpatrick {\it et al.} \cite{Kirkpatrick83} which is widely applied in the field of combinatorial optimization and which is inspired by the annealing of solids. It consists in a Metropolis procedure in which the temperature is gradually decreased. In the case of the off-line AdWords problem it is described in Algorithm \ref{algo:SA}. The inputs are an instance of the problem $\mathcal I$, an increasing sequence of inverse temperatures $\{\beta_1, \cdots, \beta_n \}$ with $\beta_1 \ll 1$, $\beta_n \gg 1$ and $n \gg 1$, and a large positive integer $T \gg 1$.

\begin{algorithm}
  \begin{pseudocode}[shadowbox]{Simulated Annealing}{\mathcal I, \{\beta_i\}, T}
  \label{algo:SA}
  \FOREACH k\in K \DO
  \BEGIN
    \mbox{select uniformly at random } a \in \partial k \\
    \FOREACH b \in \partial k \DO
      x_{kb}^0 \GETS \delta_{a,b}
  \END \\
  x_{\min} \GETS x^0 , \ E_{\min} \GETS E(x_{\min}) \\
  t \GETS 0 , \ i \GETS 1 \\
  \WHILE (i < n) \DO
    \BEGIN
    \WHILE (t < i T) \DO
    \BEGIN
      x' \GETS x^t \\
      \mbox{select uniformly at random } k \in K \\
      \mbox{select uniformly at random } b \in \partial k \setminus a \mbox{ with } a: x^t_{ka}=1 \\
      x'_{ka} \GETS 0 ,\ x'_{kb} \GETS 1 \\
      \Delta E \GETS E(x') - E(x^t) \\
      P \GETS \min [1, \, \exp(-\beta_i \Delta E)] \\
      \mbox{select uniformly at random } p \in [0,1] \\
      \IF (p < P) \THEN
        \BEGIN
          x^{t+1} \GETS x' \\
          \IF (E(x^{t+1}) < E_{\min}) \THEN
            x_{\min} \GETS x^{t+1} , \ E_{\min} \GETS E(x_{\min}) \\
        \END \\
      \ELSE
        x^{t+1} \GETS x^t \\
      t \GETS t + 1 \\
    \END \\
  i \GETS i + 1
  \END \\
  \RETURN{x_{\min}}
  \end{pseudocode}
\end{algorithm}

The values of $T$ and of $\{\beta_i\}$ must be such that the system thermalizes at any temperature $\beta_i$. In the process we can keep track of the configuration with the smallest energy explored in the sampling process, which would be the best approximation for the optimization problem we want to solve.
Additionally, keeping track of the average energy for different values of temperature, it is possible to obtain the entropy of the system as $S = \int \mathrm d E \, \beta(E)$.

\subsubsection{Linear programming relaxation with Random Rounding (RR)}

Linear programming (LP) is a common technique for optimization of linear functions over continuous variables. The variables are restricted to a given domain, usually called the {\it feasible set}, which is specified by a set of constraints that must be satisfied simultaneously by all the variables \cite{Matousek07}. Even though LP is defined on continuous variables, it is possible to exploit it in the design of approximation algorithms for problems in combinatorial optimization. The heuristics works as follows: first {\it relax} all the discrete variables in the original combinatorial problem by allowing them to take continuous values. Then use LP to find an optimal set of values for the continuous variables. In general, the optimal value for the cost function of the {\it relaxed} problem yields a lower (upper) bound for the original minimization (maximization) combinatorial problem, since the search space is enlarged by the relaxation. Moreover, if it turns out that the solution of the relaxed problem corresponds to {\it all} the variables taking integer values, then this is a solution for the discrete problem too. Otherwise, a strategy for rounding the continuous variables to integers must be specified. In general, the solution for the relaxation yields a mixture of discrete and fractional variables; it is important to bear in mind that the discrete values obtained in the solution of the LP relaxation do not necesarily correspond to an optimal solution of the original problem. Hence, in general, the rounding scheme should take into account not only the variables with fractional values, but the whole set of them instead. Figure \ref{fig:LP_failure} describes an example of integer variables that can be assigned by LP to the relaxation of the AdWords problem which are incompatible with any optimal allocation.

By introducing ancillary {\it discount} variables $\{d_a\}$ representing the discount $d_a$ given by the auctioneer to advertiser $a$ in case the actual assignment exceeds the budget $B_a$, the AdWords problem can be restated as follows
\begin{equation}
 \max\sum_{a}\left(\sum_{k} x_{ka} w_{ka} - d_a\right),\label{eq:max}
\end{equation}
subject to 
\begin{eqnarray}
 && \sum_{a} x_{ka} \leq 1 \; , \label{eq:matching_constraint}\\
 && \sum_{k} x_{ka} w_{ka} - d_a \leq B_a \; , \label{eq:budget_constraint}\\
 && x_{ka} \in \{0, 1\} \; , \label{eq:integer_constraint}
\end{eqnarray}
which is now a {\it mixed} integer programming problem, as the discount variables are allowed to take any continuous value. With this formulation, the LP relaxation consists in replacing the integer constraints in \eqref{eq:integer_constraint} with
\begin{equation}
 0 \leq x_{ka}^r,
\end{equation}
where the superscript $r$ stands for {\it relaxation}. This inequality, along with \eqref{eq:matching_constraint}, guarantees that $x_{ka}^r\leq 1$. Moreover, the discount variables $d_a$ are irrelevant in the relaxed version because the budget constraints \eqref{eq:budget_constraint} can be met by tuning the continuous variables $x_{ka}^r$ themselves; in fact they all take value zero in the solution attained by LP. They will play an important role, though, in the rounding process as described below.

A simple way to round the fractional solutions $x_{ka}^r$ given by the LP relaxation is the following \emph{Random Rounding} (RR): assign each keyword $k$ at random to advertiser $a$ with probability proportional to $x_{ka}^r$. Theorems 4 and 5 of \cite{Andelman04} prove that the approximation ratio of RR is $\mathrm e/(\mathrm e-1) \simeq 1.582$. This means that if $R_0$ is the optimal revenue for the problem considered, the solution given by RR will have a revenue $R$ satisfying the inequality
\begin{equation}
  R < R_0 < \frac{\mathrm e}{\mathrm e - 1} R \; . \label{RR_bound}
\end{equation}

\subsubsection{Linear programming with Garg-Kumar-Pandit (GKP) rounding}

Garg, Kumar and Pandit \cite{Garg01} proposed another rounding algorithm for the solution of the LP relaxation of the problem. The GKP algorithm takes the integer values obtained by LP as part of the solution, and develops a strategy to round up the remaining variables with fractional values. Theorem 3 in \cite{Garg01} proves that the algorithm to be described has an approximation ratio of $(1+\sqrt{5})/2 \simeq 1.62$, i.e.
\begin{equation}
R < R_0 < \frac{1+\sqrt{5}}{2}  R \; , \label{GKP_bound}
\end{equation}
with the same notation as in (\ref{RR_bound}). Notice that even though this bound is less favorable than (\ref{RR_bound}), the average case performance of GKP on some distribution of instances could actually be better than RR.

The rounding part of the algorithm is best described by introducing the {\it residual graph}, which is the graph obtained by removing from the original one all the edges whose associated variables were assigned an integer value by LP. The algorithm applies a series of transformations on the residual graph, consisting of a suitable redistribution of the {\it mass} on the fractional variables. In this way, without decreasing the degree of optimality reached nor violating any constraint, the residual graph is left with certain special properties; among other things, it does not contain cycles and, most importantly, it is guaranteed that any maximal path\footnote{A \emph{maximal path} is a path that cannot be extended because there are no more edges in the graph contiguous to it.} in it starts with an advertiser whose budget is saturated. If the (fractional) amount $x_{ka}^r w_{ka}$ spent by such an advertiser $a$ on the only keyword $k$ linked to him in the residual graph is less than a given fraction $p$ of the budget, i.e. $x_{ka}^r w_{ka} < pB_a$, the corresponding variable is rounded to zero; otherwise the corresponding discount variable $d_a$ is increased, opening up the possibility for the advertiser to acquire such a keyword during the remaining process \cite{Garg01}. Even though the rigorous bound (\ref{GKP_bound}) is proved for $p=(3-\sqrt{5})/2 \simeq 0.38$, for the ensemble studied here better results were observed by using smaller values of $p$.

\subsubsection{Linear programming with BP-based rounding}

Another alternative to round up the fractional variables is to use the BP algorithm to calculate the minimum energy, under the constraint that the integer variables of the LP relaxation are forced to take the actual integer values given by LP.

\begin{figure}
\input{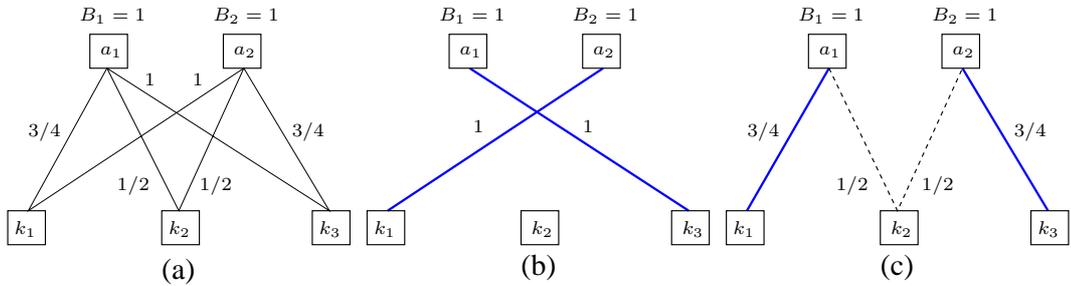}
\caption{ An example of integer variables that can be assigned by LP to the {\it relaxed} problem which are not compatible with any optimal solution: (a) Two advertisers bidding for three keywords with budget $B_a = 1$ each; advertiser $a_1$ bids $3/4$, $1/2$ and $1$ for keywords $k_1$, $k_2$ and $k_3$, respectively, while advertiser $a_2$ makes the same bids but in reverse order. (b) An {\it integer} optimal allocation with the budgets of both advertisers saturated; the missing links correspond to variables assigned to zero, except for keyword $k_2$ which can be assigned to whichever advertiser with no change in the total revenue. (c) A {\it fractional} optimal allocation, where the thick blue links correspond to the variables assigned to one, the missing links to those assigned to zero, and the dashed links to the fractional variables, which are equal to $1/2$ in this case. Notice that keyword $k_2$ is needed in order to saturate the budgets of both advertisers.} \label{fig:LP_failure}
\end{figure}

\subsubsection{Numerical results}

\begin{figure}
 \input{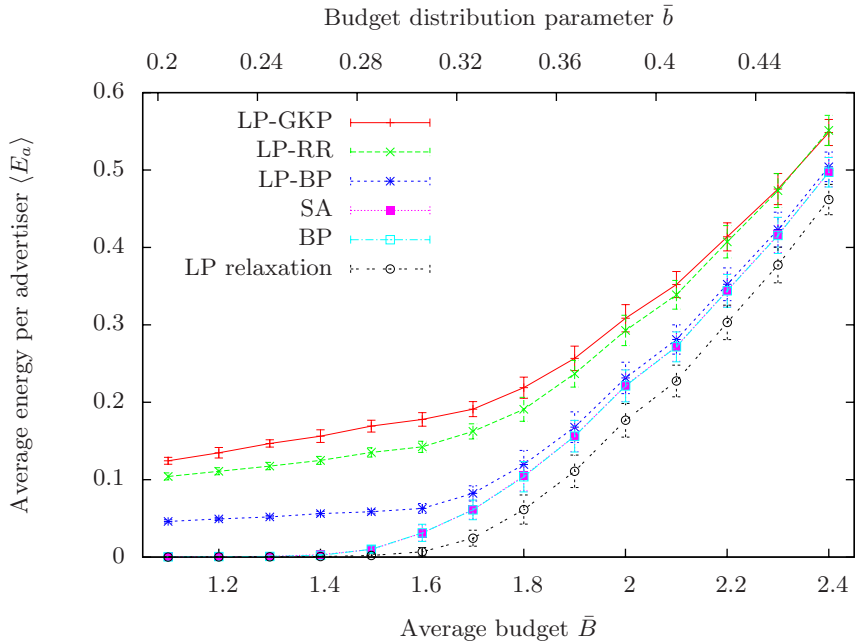}
  \caption{Comparison of the average energy per advertiser $\left< E_a \right>$ obtained by the zero-temperature BP algorithm with those obtained by alternative optimization techniques, namely: LP with random rounding (LP-RR), LP with Garg-Kumar-Pandit (LP-GKP) rounding, LP with BP-guided rounding (LP-BP), and simulated annealing (SA). The solution of the relaxed linear program provides a rigorous lower bound for the energy. The curves for SA and BP almost coincide in all the range. The same set of instances with $N_\mathrm a = 1\,000$ advertisers, $N_\mathrm k = 3\,000$ keywords, and average keyword connectivity $c_\mathrm k = 10/3$ (which yields on average $\bar N_\mathrm e = 10\,000$ variables) has been used for all the algorithms.}
  \label{comparison_plot}
\end{figure}

In order to compare these algorithms to zero-temperature BP, we have generated 300 instances from the distribution $\mathscr E(N_\mathrm a, N_\mathrm k, N_\mathrm e, \bar b)$ with $N_\mathrm a = 1\,000$, $N_\mathrm k = 3\,000$ and $N_\mathrm e \simeq 10\,000$ and with different values of the budget distribution parameter $\bar b$ and we have solved each of them with all the algorithms cosidered. We have grouped the instances in bins according to the value of the average budget $\bar B$, and we have computed the average energy of the solutions found by each algorithm in each bin. The results of the comparison are shown in Figure \ref{comparison_plot}.

The performance of BP and SA relative to the algorithms based on LP rounding appears clearly superior. In particular, notice that the performance of BP-based rounding is much worse than that of BP (especially for low budgets): the solution of the LP-relaxation typically contains some \emph{integer} variables that have the ``wrong'' value for the integer problem. An example of this can be seen in Figure \ref{fig:LP_failure}. This suggests that \emph{any} LP-rounding algorithm which assumes that the integer variables in the solution of the relaxation are correct will find a sub-optimal solution to the integer problem. This situation is very different compared to what happens for random b-matchings, a similar problem where LP and BP both give exact solutions \cite{Bayati08}.

The differences in the average energies obtained by BP and SA are always very close to zero, and since the two algorithms are completely different, this suggests that they are also very close to the true optimum. This conjecture is also supported by the analysis of the results for the entropy (not shown): as the equilibration time is increased, the entropy obtained by SA gets closer and closer to the value obtained by BP, suggesting that the latter is in fact exact, which in turn implies that the energy is exact too. Finally, the results of population dynamics for $\bar B \notin [1.41, 1.75]$, which gives the average value (over the distribution $\mathscr E_\infty$) of the actual optimal energy per advertiser, are also in excellent agreement with the energy obtained with BP and SA, confirming that they are exact at least in this range of values of $\bar B$.

It is worth noting that even though the performances of BP and SA are very similar in terms of the energy of the configurations they obtain, the two algorithms are very different in terms of running time. The running time that SA requires to find a solution with energy $E_{\mathrm SA} \leq (1+\epsilon) E_{\mathrm BP}$ is up to 100 times larger than the running time of BP for $N = 1\,000$ and $\epsilon = 0.01$.

\section{Conclusions and perspectives}

We have described a simple and efficient algorithm which is capable of finding (nearly) optimal assignments for random instances of auctions with budget constraints, based on a statistical mechanics analysis of the problem, and which is equivalent to the zero-temperature Belief Propagation procedure. We have characterized its average performance on a natural distribution of random instances and compared it to other previously studied algorithms, obtaining that it is superior to them in terms of the approximation to the true optimum and/or in terms of running time. We have also studied the properties of the instances distribution in the infinite size limit, deriving a phase diagram (as a function of the parameter $\bar b$ of the distribution) with a replica symmetry breaking phase separating two replica symmetric phases, and obtaining a reference for the average value of the true optimum (in the RS phase) against which to test our algorithm.

In the future it would be extremely interesting to test our algorithm on real data (for example of actual bids for internet search keyword auctions), or on a distribution of random instances resembling the actual one more closely. It would also be interesting to study the generalization of this off-line algorithm to the on-line problem, again from the point of view of the average performance on some distribution of search sequences (possibly time-dependent).

\subsection*{Acknowledgements}

We are grateful to C. Mathieu for introducing the problem to one of us, and to M.~M\'ezard, G.~Parisi, A.~Ramezanpour and V. Pandit for stimulating discussions. This work was partially supported by the Microsoft External Research Initiative.

\end{document}